\documentclass[conference]{IEEEtran}
\IEEEoverridecommandlockouts
\usepackage{cite}
\usepackage{amsmath,amssymb,amsfonts}
\usepackage{algorithmic}
\usepackage{graphicx}

\usepackage{stackengine}
\usepackage{comment}
\usepackage{ulem}
\usepackage{tabularx}
\usepackage{bm}

\usepackage{textcomp}
\usepackage{xcolor}
\usepackage[ruled]{algorithm2e}

\def\BibTeX{{\rm B\kern-.05em{\sc i\kern-.025em b}\kern-.08em
    T\kern-.1667em\lower.7ex\hbox{E}\kern-.125emX}}
    
\usepackage{xspace}
\usepackage{ifthen}
\usepackage[inline]{enumitem}

\newboolean{showcomments}
\setboolean{showcomments}{true}
\ifthenelse{\boolean{showcomments}}
{
}

\usepackage{subfig}
\usepackage{fancyhdr}
\begin{document}

\title{Competitive Multi-Agent Load Balancing with Adaptive Policies in Wireless Networks  \\

}

\author{\IEEEauthorblockN{ Pedro Enrique Iturria-Rivera, \IEEEmembership{Student Member,~IEEE}, Melike Erol-Kantarci, \IEEEmembership{Senior Member,~IEEE}}
\IEEEauthorblockA{\textit{School of Electrical Engineering and Computer Science}
\textit{University of Ottawa}\\
Ottawa, Canada \\
\{pitur008, melike.erolkantarci\}@uottawa.ca}}

\maketitle

\begin{abstract}
Using Machine Learning (ML) techniques for the next generation wireless networks  have shown promising results in the recent years, due to high learning and adaptation capability of ML algorithms. More specifically, ML techniques have been used for load balancing in Self-Organizing Networks (SON). In the context of load balancing and ML, several studies propose network management automation (NMA) from the perspective of a single and centralized agent. However, a single agent domain does not consider the interaction among the agents. In this paper, we propose a more realistic load balancing approach using novel Multi-Agent Deep Deterministic Policy Gradient with Adaptive Policies (MADDPG-AP) scheme that considers throughput, resource block utilization and latency in the network. We compare our proposal with a single-agent RL algorithm named Clipped Double Q-Learning (CDQL) . Simulation results reveal a significant improvement in latency, packet loss ratio and convergence time. 
\end{abstract}

\begin{IEEEkeywords}
load balancing, multi-agent deep deterministic policy gradient, self-supervision, wireless networks 
\end{IEEEkeywords}

\section{Introduction}
An undeniable growth of connected devices is experienced in the era of 5$^{th}$ Generation (5G), which is expected to continue in the future 5G and beyond networks. Besides, tremendous volume and vast diversity in high data traffic applications with stringent QoS (Quality of Service) requirements are posing a challenge for these next generation wireless networks. 

Machine learning techniques, such as Reinforcement Learning (RL), have raised a strong interest among the research community and the industry for their use in performance enhancement of wireless networks. The similarity of RL techniques with the ``trial and error'' human-like behavior conveys perfectly their usage in dynamic environments that seek ``near'' optimal control. Two main RL taxonomies are present in the literature: single agent RL and multi-agent RL. In single agent domains, an agent is the decision instance that oversees and controls the state of the environment \cite{Gronauer2021}. Although single-agent RL has shown successful performance in many problems, they might not be sufficient to fulfill the expectation in terms of reliability, latency, and efficiency, specially in wireless networks where either competition for a resource or coordination among a set of agents is essential \cite{Feriani2021}. Therefore a more realistic approach requires considering the interaction between multiple agents with the environment such as in multi-agent RL \cite{Lowe2017,Rivera2021}. 

AI-driven intelligent self-optimization functionalities are supported by 5G in which load balancing falls under the umbrella of applications considered in Self-Organizing Networks (SON). Load balancing in the context of radio access refers to the handover of UEs (User Equipments) of high loaded BSs (Base Stations) to less loaded BSs. To perform such action, the resource block utilization (RBU) per BS is observed. The amount of RBU will be related not solely on the amount of UEs attached to a specific BS but also to the traffic characteristics. One of the most known and used handover algorithms named A3 is defined in \cite{3GPP2018}. This release specifies the required events in the BS that must occur to perform a handover procedure. More specifically, it requires that neighbour BS Reference Signal Received Power (RSRP) becomes better than serving BS and this condition must hold during a predefined time named Time to Trigger (TTT) to avoid ping-pong scenarios.

\begin{figure*}
\center
  \includegraphics[scale=0.3]{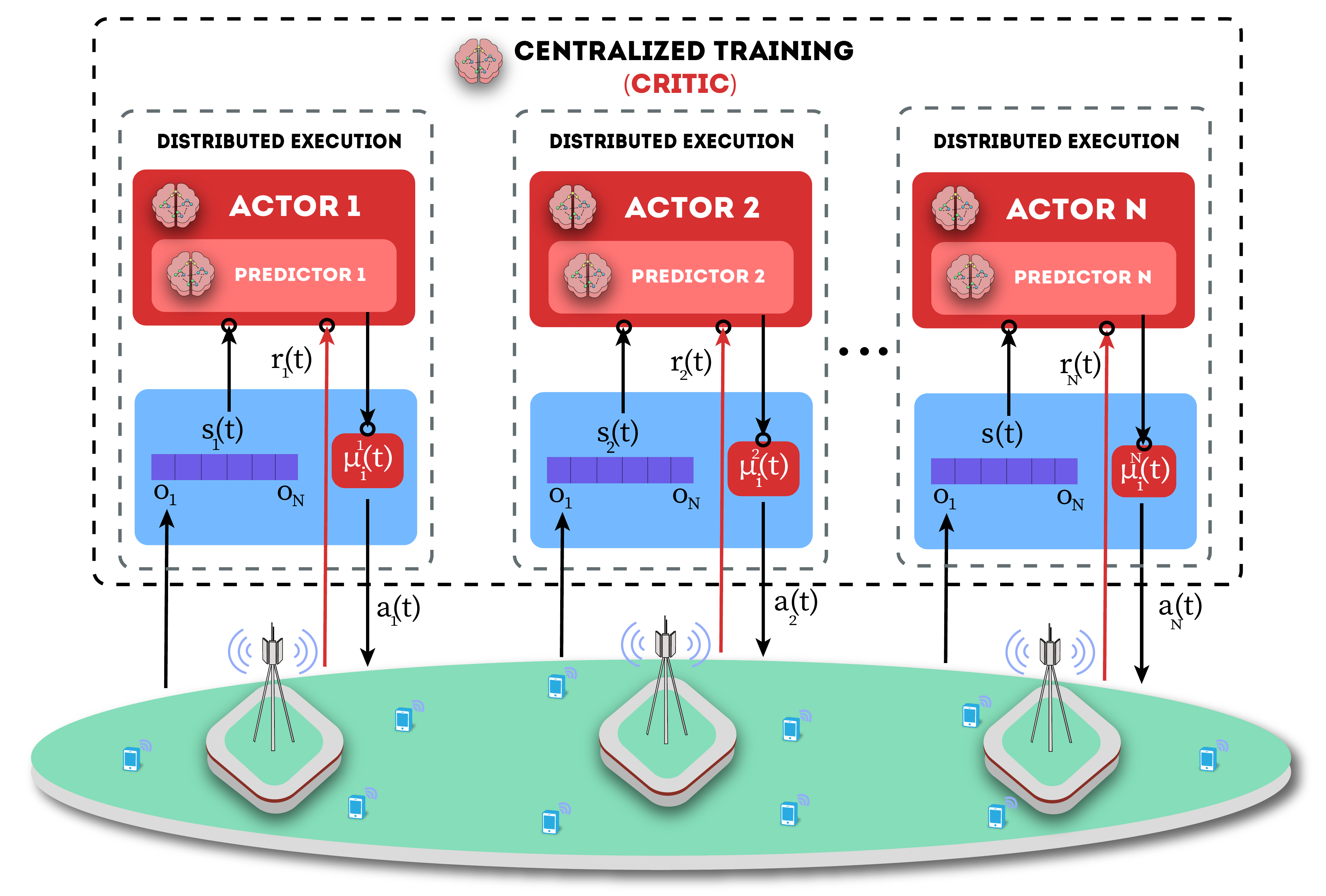}
  \caption{Multi-Agent Deep Deterministic Policy Gradient with Adaptive Policies scheme applied for load balancing in RAN. Each BS uses a sequence of past observations from the environment (state space) and utilize it as input of the policy predictor that outputs a policy and consequently the action to take. In this context, observations correspond to the attached UE's ratio, the resource block utilization and the CIO currently applied whereas the action yields the CIO value to apply.}\label{scheme}
\end{figure*}

In this paper, we address the load balancing problem in wireless networks as a competitive multi-agent problem. In a competitive setting, each agent tries to maximize its received reward under the worst-case assumption; meanwhile, the other agents always try to minimize the reward of others. Each agent represented by a BS will seek to maximize its throughput and simultaneously maintain latency and RBU constrains. 
To do so, we use a novel Multi-Agent Deep Deterministic Policy Gradient with Adaptive Policies (MADDPG-AP) algorithm that speeds up the execution stage of such centralized training-decentralized execution (CTDE) off-policy approach depicted in figure \ref{scheme}. Each agent will explore BS individual offset (CIO) values in order to maximize its objective function. We compare our results with a previous algorithm in \cite{Iturria-Rivera2021} named Clipped Double Q-Learning (CDQL) that proved its efficiency in comparison to the traditional A3 algorithm and a resource block utilization-based algorithm. Our results show an improvement in terms of latency with a 25.7\%, packet loss ratio with 28.11\% and convergence time with 70.49\% with respect to CDQL scheme. No significant improvement was obtained in terms of throughput.

\par This paper is organized as follows. Section II presents literature related to load balancing in wireless networks. In section III, the description of the system model and the motivation of this paper is introduced. Section IV provides a description of the proposed scheme. Section V depicts the performance evaluation and comparison with the baseline algorithms. Finally, section VI concludes the paper.

\section{Related work}
To the best of our knowledge there is no previous work that has dealt with load balancing in RAN from the competitive multi-agent off-policy learning perspective. However, multi-agent settings have been studied in other load-balancing approaches. Additionally, several single and centralized RL load balancing studies can be found in the literature where HO parameters are modified on the basis of network KPIs. These studies are summarized below.

In \cite{Choi2021}, the authors describe a message passing-based cooperative multi-agent Q-learning algorithm to obtain the optimal bias offset in dense HetNets (Heterogeneous
Networks). In \cite{Mai2020}, the authors study load balancing in a wired network topology with the goal of evenly distribute traffic among the nodes without incurring into congestion. To do so, the authors propose a multi-agent actor-critic to address traffic optimization.  Additionally, in \cite{Alsuhli2021}, the authors propose a RL-load balancing algorithm based on the exploration of the BS's transmission power and the CIO parameter. In this work the authors use Double Deep Q-Learning to maximize the reward function based on throughput. Finally, in \cite{Iturria-Rivera2021}, we have performed load balancing by considering resource block utilization, delay and CQI metrics in its reward objective function. This work is different than our previous work in terms of the consideration of a multi-agent domain with an adaptive and continuous reinforcement learning scheme and a new formulation of the Markov Decision Process (MDP).  

\section{System model} 
\label{AA}
We consider a network consisting of $M_T$ base stations (BS) where for each $i_{th}$ BS, $i \in \Lambda$. The network serves a set of $\Psi$ of size $N_T$ stationary mobile users randomly deployed around the BSs. The Channel and QoS Aware (CQA) scheduler is chosen as MAC scheduler.In this paper, we consider one Resource Block Group (RBG) ($RBG = N_{RB}^{DL} / K$) as the smallest resource unit where $K\in[1,2,3,4]$ according the DL bandwidth used. Additionally, each BS is considered an agent of our environment. Each agent is capable of modifying their own CIO value based on the exploration performed by our RL scheme. The CIO value is utilized as part of the A3 handover algorithm that will indicate handover between BS $i$ and BS $j$ if the following condition holds: $RSPR_j + \o_{j\rightarrow i} > Hys + RSPR_i + \o_{i\rightarrow j}$. $RSPR_j$ and $RSPR_i$ are the measured RSRP (Reference Signal Received Power) values in dB of the serving BS and the neighbor BS, respectively. $Hys$ is the hysteresis value used to avoid ping-pong scenarios.

\section{Multi-Agent Deep Deterministic Policy Gradient with Adaptive Policies}
In the proposed approach, each agent observes attached UEs’ ratio, the resource block utilization and the currently chosen CIO value and receives its reward based on the feedback from the environment in terms of as throughput, packet delay and resource block utilization. The agent's actions consist of modifying the BS's individual CIO values in order to maximize the agent's reward. In the following subsection, we present an overview of MADDPG-AP and then formally define our solution.

\subsection{Multi-agent Deep Deterministic Policy Gradient }\label{AA}

In this work, we use a state-of-the-art policy gradient algorithm. This algorithm is presented in \cite{Lowe2017} and builds on the generalization for the multi-agent domain of Deep Deterministic Policy Gradient algorithm (DDPG) \cite{Lillicrap2016}. MADDPG is a multi-agent off-policy and continuous action space algorithm constrained to three main assumptions: (1) the learned policy per agent uses local observations, (2) the environment is non-deterministic, thus a differentiable model is not assumed and (3) a communication structure is not assumed among agents. Furthermore, this algorithm belongs to the centralized training with decentralized execution family.

\subsection{Addressing non-stationarity among competitive agents in MADDPG }\label{AA}
Non-stationarity, mainly caused by the joint interaction of the agents with the environment, is one of the primary issues presented by the MARL setting. When the environment experiments a non-stationary behavior it becomes a moving target problem where each agent's best policy changes as individual policies change\cite{Hernandez-Leal2019}. The authors in \cite{Lowe2017} called the attention to the previously mentioned issue in the particular case of competitive agents due the natural overfitting of a single agent's behavior over its competitors. To address this problem, it was proposed training an ensemble of $K$ sub-policies per agent where each agent had to choose a random $k$ sub-policy per episode. Another proposed solution, was M3DDPG (MiniMax Multi-agent Deep Deterministic Policy Gradient) \cite{Li2019}, inspired by MADDPG, where the authors consider a minimax approach to train robust agents even when the opponent agent's performance is considered not good. Finally, in \cite{WWprima20} the authors consider that by having $K$ learned subpolicies per agent under the same constrains as MADDPG, the agent would be able to choose a corresponding policy according to the state observed. Our algorithm is mainly based on \cite{WWprima20} saving some particular differences that will be addressed in subsection C.

\subsection{Adaptive policies and ranked buffer}

\begin{figure}
\center
  \includegraphics[scale=0.6]{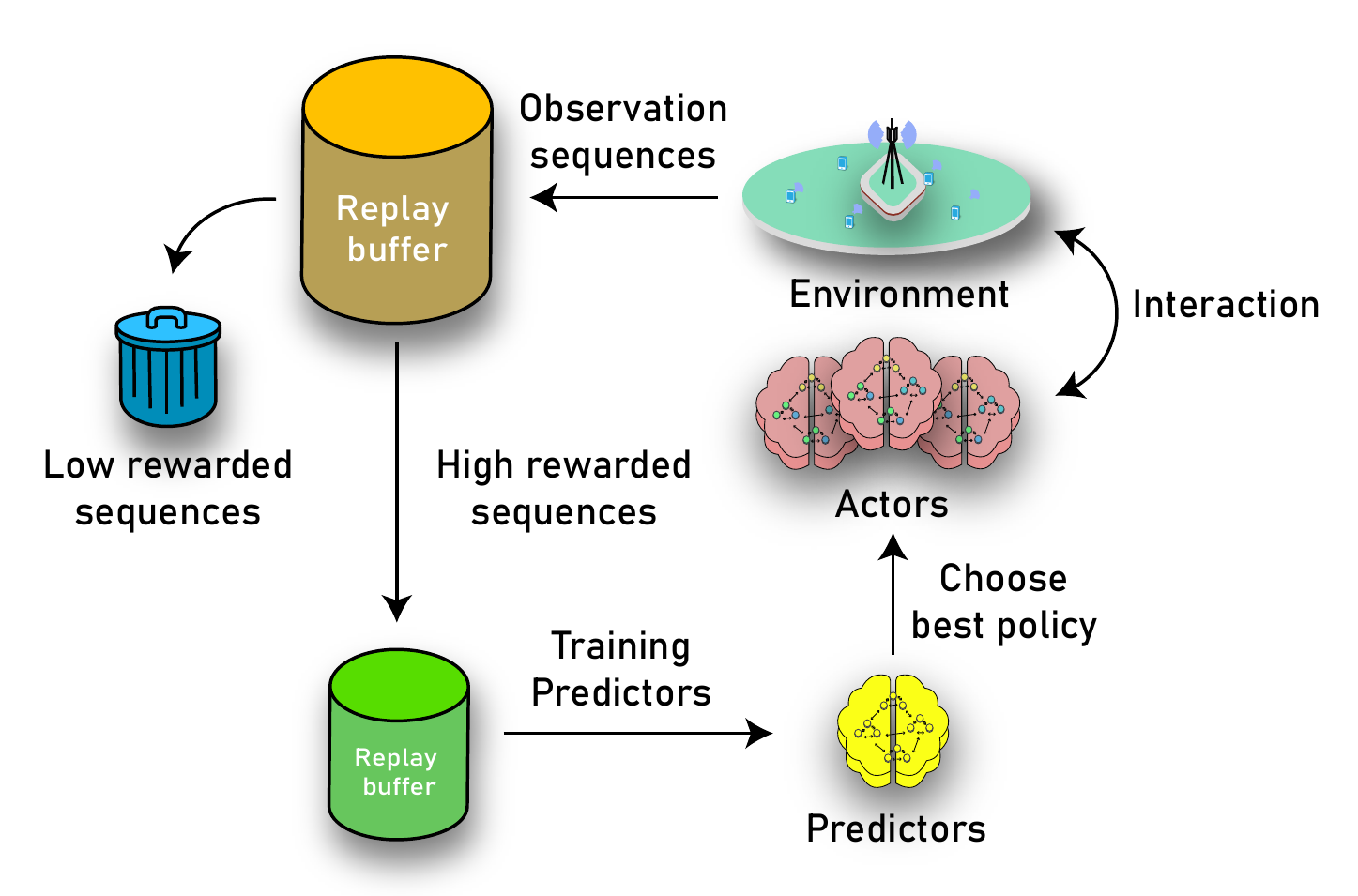}
  \caption{Ranked buffer in MADDPG-AP} 
  \label{rank_buffer}
\end{figure}

The curse of dimensionality and collecting data during training are the main reasons why a slow exploration is experienced in off-policy MARL algorithms \cite{Ye}. For instance, in the context of the RIC (RAN Intelligent Controller) in the O-RAN architecture (Open RAN) there are some close loop requirements in terms of latency that algorithms must comply \cite{Iturria-Rivera2021}. Thus, a reasonable time must be considered executing any models. To accelerate such convergence time and inspired by \cite{Zha2021}, we introduce a self-supervised technique using a ranked buffer in the policy predictors' training stage. As a main difference,  we rank state sequences, instead of local observations, by the sum of the reward observed during such sequence and discard those sequences with low reward (see Algorithm \ref{rank_buffer}) to form a new buffer that is used to train the the policy predictor as seen in Figure \ref{rank_buffer}. By doing the latter, we select the best subpolicy given an observed sequence. 

\normalem 
\begin{algorithm}

\algsetup{linenosize=\tiny}
 \scriptsize

\SetAlgoLined

\ForEach{agent $i$ to $N$} {

Create sequences from $\mathcal{B}$ $\bm{w}_i = <(o_1,..,o_W)_1,...,(o_1,..,o_W)_{N_s}>$ using a sliding window of size $W$ \\
Rank sequences in $\bm{w}_h$ based on $\hat{r}_{N_s} = \sum_{i=0}^{W}r_i$ and obtain a new buffer $\hat{\mathcal{B}}$ and $B_s = N_s/K$.\\
\If{$|\hat{\mathcal{B}}| > B_s $} {
Discard sequences with low reward in $\hat{r}_{N_s}$.
}
}
     \label{rank_buffer}
    
     \caption{Ranked buffer}
 \medskip
\end{algorithm}

\normalem 
\begin{algorithm}
\algsetup{linenosize=\tiny}
 \scriptsize

\SetAlgoLined
{\fontfamily{pcr}\selectfont
\# Training stage:
}\\
\ForEach{agent $i$ to $N$} {
    Init  set of policies $\Pi_i^K = <\mu_i^1,...,\mu_i^K>$ and predictors   $\rho_i$ \\
    \For{$k\gets0$ subpolicy \textbf{to} $K$} {
    Learn $\Pi_i^k$ and $\rho_i$
}
}

{\fontfamily{pcr}\selectfont
\# Execution stage:} \\
Load predictors $\rho_i$ and set of policies $\Pi_i^K$ per agent, respectively.

\For{environment step $t\gets1$ \textbf{to} $T$}{ 
    \ForEach{agent $i$ to $N$} {
    $o_i^t \leftarrow$ receive observation and append it to $\bm{w}_i^t$\\
    $\mu_i^k \leftarrow$ predict policy by $\rho_i(\bm{w}_i^t)$; $|\bm{w}_i^t|= W$ \\
    $a_i^t \leftarrow$ select action by  $\mu_i^k$\\
}
    Execute actions $\bm{a} = (a_1^t,...,a_N^t)$\\
    Collect rewards $\bm{r} = (r_1^t,...,r_N^t)$\\
} 

     \label{amaddpg_desc}
    
     \caption{Training and execution stages for AMADDPG algorithm}
 \medskip
\end{algorithm}

The pseudo-code for AMADDPG with ranked buffer is presented in Algorithm \ref{amaddpg_desc} whereas the training stage of AMADDPG is depicted in Algorithm \ref{amaddpg_training}.

\subsection{Action space selection}\label{AA}
The action value of each agent corresponds to the CIO value selected in each timestep and is defined as a continuous variable limited by a predefined exploration bound. Thus,
\begin{equation}
    A_j(t) = \o_j(t) 
\end{equation}

The individual CIO value is lower and upper bounded by two predefined values $\o_{min}$ and $\o_{max}$, respectively, as described as follows: $\o_{i}(t) \in [\o_{min}, \o_{max}]$. Furthermore, Ornstein–Uhlenbeck noise is used to encourage exploration among the agents.

\normalem 
\begin{algorithm}

\algsetup{linenosize=\tiny}
\scriptsize

\SetAlgoLined
 Learning stage: Init $N$ predictors networks $\rho_i$  

 \ForEach{K subpolicy}{
 Initialize $N$ actors and target networks $\mu_{i}$ and $\mu_{i}'$ respectively, the centralized critic  $Q_{i}^{\mu}$ and target critic network with bounded random weights  and replay buffer $\mathcal{B}$ \;
 \ForEach{episode step}{
	    Initialize $N$ Ornstein–Uhlenbeck random processes for action exploration\; 
	   \ForEach{environment step}{
	        for each agent $i$, select action $a_i=\mu_{\theta_i} + \mathcal{N}_t$ w.r.t the current policy and exploration\;
	        Execute actions $\textbf{a} = (a_1,...,a_N)$ \;
	        Store $(\textbf{x},\textbf{a},\textbf{r},\textbf{x}')$ in replay buffer $\mathcal{B}$ \;
	        $\textbf{x} \leftarrow \textbf{x}'$ \;
	        \ForEach{agent $i$ to $N$}{
	           Sample a random minibatch of $S$ samples $(\textbf{x}^j,\textbf{a}^j,\textbf{r}^j,\textbf{x}^{j\prime})$ from $\mathcal{B}$\;
	           Set $y^j = r_i^j + \gamma Q_i^{\bm{\mu}'}(\textbf{x}^{j\prime},a'_1,...,a'_N)|_{a'_k = \bm{\mu}'_k(o_k^j)}$\\
	           Update actor $i$ by using the sampled policy gradient:\\
	           $\nabla_{\theta_i}J \approx \frac{1}{S}\sum_{s=1}^{S}  \nabla_{\theta_i}\bm{\mu}_i(o_i^j)$\\ $\nabla_{a_i}Q_i^{\bm{\mu}}(\textbf{x}^j, a_i^j,...,a_i,...,a_N^j)|_{a_i=\bm{\mu}_i(o_i^j)}$\\
	           Update critic by minimizing the loss:\\
	           $\mathcal{L}(\theta_i)=\frac{1}{S}\sum_{s=1}^{S} (y^j - Q_i^{\bm{\mu}}(\textbf{x}^j,a_i^j,...,a_N^j))^2$
	        }
	        Update target networks parameters\;
	        $\theta_{i}' \leftarrow \tau *\theta_{i} + (1-\tau) * \theta_{i}'$ \;
       
    	}

    }

	Sample training data $K$ from $\hat{\mathcal{B}}$ and train LSTM predictor $\rho_i$ networks with sequences as input and the corresponding subpolicy as output by minimizing the following negative log likelihood loss:\\
	
	\ForEach{agent $i$ to $N$} {
	            $\nabla_{\rho_i}J(\rho_i) \approx$ $\frac{1}{K}\sum_{k=1}^{K}\sum_{\mu'_k}-y^{\mu'_i}log(p_i(\mu'_i))$
	 }
 }
\label{amaddpg_training}
\caption{Multi-Agent Deep Deterministic Policy Gradient with Adaptive Policies and Ranked Buffer}

\end{algorithm}

\subsection{State space selection}\label{AA}
Each BS state space is composed by three terms. The first term corresponds to the attached UEs' ratio, the second term is the Resource Block Utilization (RBU) and lastly, the current CIO value. Thus, the state $S(t)$ will be defined by the concatenation of such metrics as: 

\begin{equation}
    S_j(t) = \begin{bmatrix} u_j(t) & p_j(t)  & \o_j(t) \end{bmatrix}
\end{equation}
where $u_j = \dfrac{\Psi_{\Gamma_{i}}}{N_T}$ and $\Psi_{\Gamma_{j}}$ is the total of UEs attached to BS $j$.

Additionally, $p(t)$ corresponds to the RBU for BS $j$ at time $t$. Resource block allocation for each BS is gathered every agent's observation time (0.2s) which is a greater than the TTI. Thus, we model the resource block utilization during the observation time $p_i$ as the expected value of the resource block utilization for each TTI as $p_{tti}$. 
\begin{equation}
    p_j = \mathbb{E}[p_{tti}]
\end{equation}

\subsection{Reward}\label{AA}
The total reward function is calculated based on Quality of Service (QoS) parameters  and the load of each BS in terms resource block utilization. Let us start with a well-known reward based on throughput\cite{Alsuhli2021}:
\begin{equation}
   \mathcal{R}_{j} = \sum_{i=0}^{U_c} R_{i}
   \label{sum_thr}
\end{equation}

where,  $U_c$ is the number of UEs connected to BS $j$ and $R_{i}$ corresponds to the throughput measured at time $t$ by UE $i$. 
This reward does not consider RBU and latency and it does not contain any information related to the UE's connection status (For example, given certain value of CIO, UE $i$ can be attached to BS $j$ but not achieve connectivity, in other words, delay is infinite or throughput is equal to 0). Our approach consists of adding an extra parameter that will either punish or reward the utility function defined in Eq. \ref{sum_thr}.
The mean throughput is chosen as defined in Eq. \ref{beta_sum}. The idea behind it is to use an approximation of the average throughput and to penalize if there are disconnected UEs. 
The second term in the sum of Eq. \ref{beta_sum} will penalize or reward the equation Eq. \ref{sum_thr}  depending on $\epsilon_1$ where $\epsilon_1$, is a latency-based parameter as defined later in Eq. \ref{eps_1} that will oversee penalizing or rewarding according the latency requirements for a specific application.
\begin{equation}
  \mathcal{R}_{j} = \sum_{i=0}^{U_c} R_{i} + \left[\frac{\epsilon_{1i}}{U_c}\sum_{i=0}^{U_c} R_{i}\right] 
   \label{beta_sum}
\end{equation}
After some algebraic modification Eq. \ref{beta_sum2}. is obtained.
\begin{equation}
   \mathcal{R}_{j} = \left[\frac{U_c + \sum_{i=0}^{U_c} \epsilon_{1i}}{U_c}\right]\sum_{i=0}^{U_c} R_{i}
   \label{beta_sum2}
\end{equation}

\begin{figure*}
\center
  \includegraphics[scale=0.65]{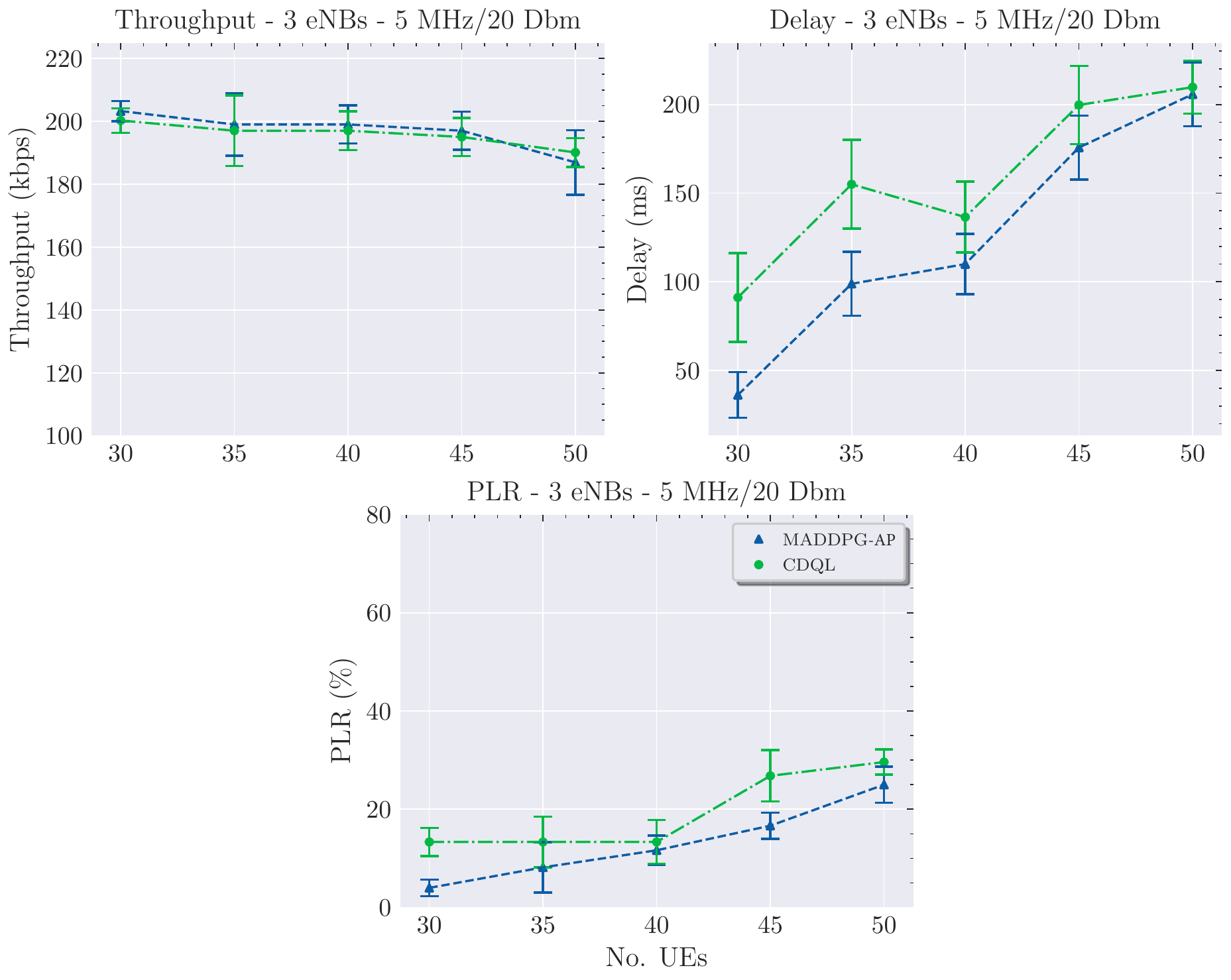}
  \caption{Performance metrics of MADDPG-AP and CDQL algorithms. (a) Throughput, (b) Delay and (c) PLR }
  \label{metrics}
\end{figure*}

The current reward formulation in Eq. \ref{beta_sum2} is composed by two KPIs that are UE-dependent. However, the RBU-based reward tries to minimize the number of resources that an individual BS is managing. Since our $\mathcal{R}_{j}$ is calculated based on BS, the reward based on RBU  $\epsilon_2$ depicted in Eq. \ref{second_rew}  is multiplied by the term in Eq. \ref{beta_sum2} obtaining: 
\begin{equation}
   \mathcal{R}_{j} = \epsilon_{2j}\left[\frac{U_c + \sum_{i=0}^{U_c} \epsilon_{1i}}{U_c}\right]\sum_{i=0}^{U_c} R_{i}
   \label{sum_final}
\end{equation}

$\epsilon_{1} \in [-1,1]$ can be defined as: 

\def\delequal{\mathrel{\ensurestackMath{\stackon[1pt]{=}{\scriptstyle\Delta}}}}

\begin{equation}
   \epsilon_{1}(t)\delequal \frac{1}{N_T} \sum_{i=1}^{N_T} \mathbb{1}\cdot\delta_{i}(t) 
    \label{eps_1}
\end{equation}
and, 
\begin{equation}
   \delta_{i}(t) =
    \begin{cases}
      -1 & \text{if $\exists \Psi_i \notin \Psi_C$,}\\
      \daleth(D_{avg}) & \textit{otherwise}
    \end{cases}       
\end{equation}
where $\Psi_i$ represents the $i$th UE and $\Psi_C$, $\Psi_C\subset \Psi$ is the set of connected UEs to any BS $\in \Gamma$. On the other hand, $\daleth(D_{avg})$ is a sigmoid function defined as follows: 

\begin{equation}
   \daleth(D_{avg}) = 1 + \dfrac{c}{1+ e^{-o(D_{avg} - \mathcal{F})}}
\end{equation}
Here $c$ establishes the upper bound of the slope, $o$ adjusts the slope of the sigmoid and $\mathcal{F}=2/3*PDB$ controls the target packet delay. $PDB$ is the Packet Delay Budget which according the type of traffic used in the network.

The objective of $\epsilon_{1}$ itself is to reward each UE's average latency based on the PDB of a defined packet type. A penalization is applied if an UE is found disconnected from the network due a load balancing decision. 

Finally, $\epsilon_{2} \in [0,1]$ can be defined as:
\begin{equation}
\epsilon_{2}(t)  \delequal 1 + \dfrac{c}{2+ 2e^{-a(p(t) - \mathcal{D})}}
\label{second_rew}
\end{equation}
where $p(t)$ corresponds to the resource block utilization measured value by the BS at time t.  As $p(t)$ decreases, higher reward is obtained. $c$ establishes the upper bound of the slope, $a$ adjusts the slope of the sigmoid as in Eq. \ref{second_rew} and $\mathcal{D}$ controls the target resource block utilization.

\vspace{-0.1cm}
\subsection{Baseline: Clipped Double Q-Learning (CDQL) }\label{AA} 
For the present work, as a baseline, we consider a single and centralized reinforcement learning algorithm named CDQL that was proposed in \cite{Iturria-Rivera2021}. This technique showed an improvement of throughput, latency, jitter and PLR when compared to traditional handover algorithms. The CDQL presents two main differences with our current work: $i)$ It considers a single and centralized agent with a deterministic action space and $ii)$ It does not consider maximizing throughput in its reward design.

\section{Performance Evaluation}
\subsection{Simulation Setting}\label{AA}

\begin{table} 
\caption{Network settings}
\begin{center}
\resizebox{\columnwidth}{!}{%
\begin{tabular}{c c} 
\hline\hline
\textbf{Parameter}&\textbf{Value} \\
\hline

BS Inter-site Distance & {720 m} \\
$M_T$ & { 3 } \\
$N_T$ & { 30,35,40,45,50 } \\
Center Frequency & {2 MHz} \\
System Bandwidth & { 5 MHz (25 resource blocks)} \\ 
Pathloss Model & { Log Distance Propagation Loss Model } \\
\textbf{} & { 95 + 27 $log_{10}(distance[km])$ } \\
BS antenna height & { 30 m } \\
UE antenna height & { 1.5 m} \\
Max Tx power & {20 dbm } \\  
MAC scheduler & { CQA scheduler} \\
User distribution & { Stationary and uniformly distributed} \\
\hline
Traffic Model & { Conversational video (live streaming) and Poisson } \\
\textbf{} & { Packet payload size = 250 Bytes } \\
\textbf{} & { Interval = 10 ms } \\
\textbf{} & { Packet delay budget = 150 ms } \\
Handover algorithm & { A3-event based } \\
 & { Time to trigger = 8 ms } \\
  & { Hysteresis =  2 dBm } \\
\hline
\end{tabular}
}
\label{net_settings}

\end{center}
\vspace{-7mm}
\end{table}

Simulations are implemented using the discrete network simulator ns-3. Additionally, OpenAI Gym is utilized to interface the ns-3 wireless environment and to our proposed algorithm. In Table \ref{net_settings} and Table \ref{q_settings} the settings used in our simulations and the RL parameters are given, respectively. The inter-site distance between BS is set to 720 meters. Five different scenarios are tested under the proposed algorithms with 30, 35, 40, 45 and 50 UEs. The UEs are distributed on the edge of each BS in a random disc and the rest is uniformly allocated throughout the coverage of the middle BS.
\vspace{-0.3cm}
\begin{table}[htbp]
\caption{Learning parameters.}
\begin{center}
\resizebox{\columnwidth}{!}{%
\begin{tabular}{c c} 
\hline\hline
\textbf{Parameter}&\textbf{Value} \\
\hline

\# Training stage
Number of iterations/episode & {50} \\
Number of episodes & { 300  } \\
\# Execution stage
Number of iterations/episode & {50} \\
Number of episodes & { 150 } \\
Gym environment step time & { 0.2s } \\
Batch size & { 100 } \\
\hline
MADDPG-AP & $\o_ \in [-9,9]$ dBm \\
\textbf{} & $K = 3$  \\
\textbf{} & $\mathcal{F}=2/3*PDB$ where $PDB= 150$ ms, $c=-2 $, $o=75$\\
\textbf{} & $\gamma_{RB} = 0.8$, $a = 20$\\
\textbf{} & Optimizer :  {Actor: Adam (1e-4), Critic: Adam (1e-3)} \\
\textbf{} & Number of hidden layers ($N_h$) : {2} \\
\textbf{} & Number of neurons/layer ($n_l$) : {128} \\
\textbf{} & Update target model type : Polyak averaging \\
\textbf{} & $\gamma=0.95, \epsilon=1.0, \epsilon_{min} = 0.001, \epsilon_{decay} = 0.995$\\
\hline 
\end{tabular}
\label{q_settings}
}
\end{center}
\vspace{-4mm}
\end{table}


\vspace{-0.6cm}

\subsection{Simulation Results}\label{AA}

We present the performance results of our proposed scheme in terms of throughput, delay, and packet loss ratio (PLR) with 90\% confidence interval. Figures \ref{metrics} (b and c) show a positive change with respect to the baseline with a gain of 20.89\% in terms of delay and 32.1\% in terms of PLR. No evident improvement was observed in terms of throughput, as seen in Figure \ref{metrics} (a).

\begin{figure}
\center
  \includegraphics[scale=0.75]{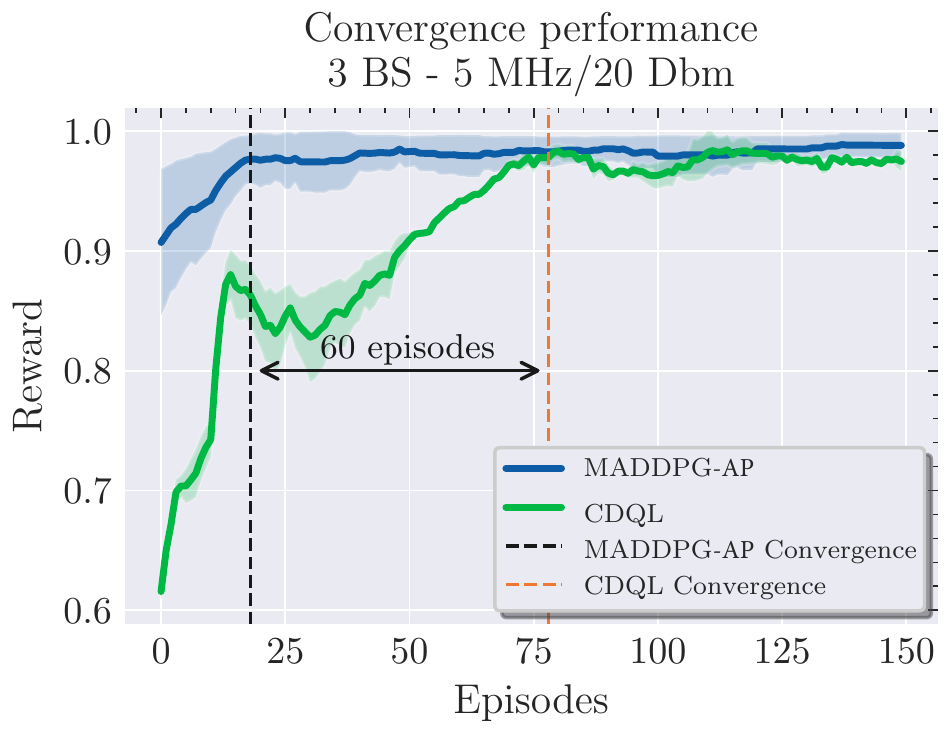}
  \caption{Convergence performance for the 30 UEs scenario for the CDQL and MADDPG-AP schemes.}
  
  \label{convergence}
\end{figure}

Figure \ref{convergence} presents a convergence comparison for the 30 UEs scenario for the MADDPG-AP and CDQL algorithms where an improvement up to 60 episodes is achieved. Finally, Table \ref{time_conv} shows an average convergence time over all scenarios with an improvement of 70.49\% for MADDPG-AP with respect to the CDQL scheme.


\vspace{-0.3cm}
\begin{table}[ht]

\caption{Convergence comparison} 
\centering 
\begin{tabular}{c c c } 
\hline\hline 
RL Scheme &Avg. Convergence in Episodes & Avg. Improvement \\ [1ex] 
\hline 
CDQL & 61 (12.2s) &   -  \\ 
MADDPG-AP & \textbf{18} (\textbf{3.6}s) & 70.49\% \\ [1ex] 
\hline 
\end{tabular}
\label{time_conv} 
\end{table}

\vspace{-0.3cm}
\section{Conclusions }
In this paper, we presented an Multi-Agent Deep Deterministic Policy Gradient with Adaptive Policies and ranked buffer strategy where BSs compete to achieve load balancing with awareness of QoS metrics. Our work differentiates from previous approaches mainly with the proposal of a state-of-the-art adaptive competitive MARL with continuous action state scheme with self-supervision capabilities. We compared our proposed method with a baseline (CDQL) utilized in a previous work that has overcome the performance of traditional handover algorithms. Although no evident improvement was obtained in terms of throughput, significant gain was obtained in terms of average delay with a 25.7\% and PLR with 28.11\%, in comparison to CDQL scheme. Additionally, an average improvement was achieved in the execution stage of 70.49\% with respect to CDQL scheme.  


\section{Acknowledgment }
This research is supported by the 5G ENCQOR program and Ciena. 

\vspace{-0.3cm}
\bibliography{biblio.bib}{}

\begin{thebibliography}{10}
\providecommand{\url}[1]{#1}
\csname url@samestyle\endcsname
\providecommand{\newblock}{\relax}
\providecommand{\bibinfo}[2]{#2}
\providecommand{\BIBentrySTDinterwordspacing}{\spaceskip=0pt\relax}
\providecommand{\BIBentryALTinterwordstretchfactor}{4}
\providecommand{\BIBentryALTinterwordspacing}{\spaceskip=\fontdimen2\font plus
\BIBentryALTinterwordstretchfactor\fontdimen3\font minus
  \fontdimen4\font\relax}
\providecommand{\BIBforeignlanguage}[2]{{%
\expandafter\ifx\csname l@#1\endcsname\relax
\typeout{** WARNING: IEEEtran.bst: No hyphenation pattern has been}%
\typeout{** loaded for the language `#1'. Using the pattern for}%
\typeout{** the default language instead.}%
\else
\language=\csname l@#1\endcsname
\fi
#2}}
\providecommand{\BIBdecl}{\relax}
\BIBdecl

\bibitem{Gronauer2021}
S.~Gronauer and K.~Diepold, ``{Multi-agent deep reinforcement learning: a
  survey},'' \emph{Artificial Intelligence Review}, 2021.

\bibitem{Feriani2021}
A.~Feriani and E.~Hossain, ``{Single and Multi-Agent Deep Reinforcement
  Learning for AI-Enabled Wireless Networks: A Tutorial},'' \emph{IEEE
  Communications Surveys and Tutorials}, 2021.

\bibitem{Lowe2017}
R.~Lowe, Y.~Wu, A.~Tamar, J.~Harb, P.~Abbeel, and I.~Mordatch, ``{Multi-agent
  actor-critic for mixed cooperative-competitive environments},'' in
  \emph{Advances in Neural Information Processing Systems}, 2017.

\bibitem{Rivera2021}
\BIBentryALTinterwordspacing
P.~E. Iturria-Rivera, S.~Mollahasani, and M.~Erol-Kantarci, ``{Multi Agent Team
  Learning in Disaggregated Virtualized Open Radio Access Networks (O-RAN)},''
  2020. [Online]. Available: \url{http://arxiv.org/abs/2012.04861}
\BIBentrySTDinterwordspacing

\bibitem{3GPP2018}
3GPP, ``{TS 36.331: Radio Resource Control (RRC); Protocol specification
  (Release 15)},'' \emph{3Gpp}, 2018.

\bibitem{Iturria-Rivera2021}
P.~E. Iturria-Rivera and M.~Erol-Kantarci, ``{QoS-Aware Load Balancing in
  Wireless Networks using Clipped Double Q-Learning},'' in \emph{2021 IEEE 18th
  International Conference on Mobile Ad-Hoc and Smart Systems, MASS 2021},
  2021.

\bibitem{Choi2021}
H.~Choi, T.~Kim, H.-s. Park, and J.~K. Choia, ``{A Cooperative Online
  Learning-Based Load Balancing Scheme for Maximizing QoS Satisfaction in Dense
  HetNets},'' \emph{IEEE Access}, 2021.

\bibitem{Mai2020}
T.~Mai, H.~Yao, Z.~Xiong, S.~Guo, and D.~T. Niyato, ``{Multi-agent Actor-Critic
  Reinforcement Learning Based In-network Load Balance},'' in \emph{2020 IEEE
  Global Communications Conference, GLOBECOM 2020 - Proceedings}, 2020.

\bibitem{Alsuhli2021}
G.~Alsuhli, H.~A. Ismail, K.~Alansary, M.~Rumman, M.~Mohamed, and K.~G. Seddik,
  ``{Deep reinforcement learning-based CIO and energy control for LTE mobility
  load balancing},'' in \emph{2021 IEEE 18th Annual Consumer Communications and
  Networking Conference, CCNC 2021}, 2021.

\bibitem{Lillicrap2016}
T.~P. Lillicrap, J.~J. Hunt, A.~Pritzel, N.~Heess, T.~Erez, Y.~Tassa,
  D.~Silver, and D.~Wierstra, ``{Continuous control with deep reinforcement
  learning},'' in \emph{4th International Conference on Learning
  Representations, ICLR 2016 - Conference Track Proceedings}, 2016.

\bibitem{Hernandez-Leal2019}
P.~Hernandez-Leal, B.~Kartal, and M.~E. Taylor, ``{A survey and critique of
  multiagent deep reinforcement learning},'' \emph{Autonomous Agents and
  Multi-Agent Systems}, 2019.

\bibitem{Li2019}
S.~Li, Y.~Wu, X.~Cui, H.~Dong, F.~Fang, and S.~Russell, ``{Robust multi-agent
  reinforcement learning via minimax deep deterministic policy gradient},'' in
  \emph{33rd AAAI Conference on Artificial Intelligence, AAAI 2019, 31st
  Innovative Applications of Artificial Intelligence Conference, IAAI 2019 and
  the 9th AAAI Symposium on Educational Advances in Artificial Intelligence,
  EAAI 2019}, 2019.

\bibitem{WWprima20}
Y.~Wang and F.~Wu, ``Policy adaptive multi-agent deep deterministic policy
  gradient,'' in \emph{Proceedings of the 23rd International Conference on
  Principles and Practice of Multi-Agent Systems (PRIMA)}, Nagoya, Japan,
  Novermber 2020.

\bibitem{Ye}
Z.~Ye, Y.~Chen, G.~Song, B.~Yang, and S.~Fan, ``{Experience Augmentation:
  Boosting and Accelerating Off-Policy Multi-Agent Reinforcement Learning}.''

\bibitem{Zha2021}
\BIBentryALTinterwordspacing
D.~Zha, K.-H. Lai, K.~Zhou, and X.~Hu, ``{Simplifying Deep Reinforcement
  Learning via Self-Supervision},'' jun 2021. [Online]. Available:
  \url{https://arxiv.org/abs/2106.05526v1}
\BIBentrySTDinterwordspacing

\end{thebibliography}
\bibliographystyle{IEEEtran}

\end{document}